\documentclass[conference,a4paper]{IEEEtran}

\usepackage{times}
\usepackage[final]{graphicx}
\usepackage[reqno]{amsmath}
\usepackage{amsfonts}
\usepackage{caption2}
\usepackage{times,amsmath,epsfig,psfrag}
\usepackage{latexsym,amssymb}
\usepackage{cite}

\renewcommand{\baselinestretch}{0.95}
\def\myQED{\mbox{\rule[0pt]{1.5ex}{1.5ex}}}

\newtheorem{thm}{Theorem}

\newtheorem{prop}[thm]{Proposition}
\newtheorem{lem}[thm]{Lemma}
\newtheorem{rmk}{Remark}

\newcommand{\no}{\nonumber}
\newcommand{\phiv}{\hbox{\boldmath$\phi$}}

\IEEEoverridecommandlockouts

\begin{document}

\sloppy

\title{Quickest Search Over Multiple Sequences with Mixed Observations}

\author{
  \IEEEauthorblockN{Jun Geng}
  \IEEEauthorblockA{Dept. of Electrical \& Computer Engr.\\
    Worcester Polytechnic Institute\\
    Email: jgeng@wpi.edu}
  \and
  \IEEEauthorblockN{Weiyu Xu}
  \IEEEauthorblockA{Dept. of Electrical \& Computer Engr.\\
    Univ. of Iowa\\
    Email: weiyu-xu@uiowa.edu}
  \and
  \IEEEauthorblockN{Lifeng Lai\thanks{The work of J. Geng and L. Lai was supported by the National Science Foundation CAREER award under grant CCF-13-18980 and by the National Science Foundation under grant DMS-12-65663.}}
  \IEEEauthorblockA{Dept. of Electrical \& Computer Engr.\\
    Worcester Polytechnic Institute\\
    Email: llai@wpi.edu}
}



\maketitle

\begin{abstract}
The problem of sequentially finding an independent and identically distributed (i.i.d.) sequence that is drawn from a probability distribution $F_1$ by searching over multiple sequences, some of which are drawn from $F_1$ and the others of which are drawn from a different distribution $F_0$, is considered. The sensor is allowed to take one observation at a time. It has been shown in a recent work that if each observation comes from one sequence, Cumulative Sum (CUSUM) test is optimal. In this paper, we propose a new approach in which each observation can be a linear combination of samples from multiple sequences. The test has two stages. In the first stage, namely scanning stage, one takes a linear combination of a pair of sequences with the hope of scanning through sequences that are unlikely to be generated from $F_1$ and quickly identifying a pair of sequences such that at least one of them is highly likely to be generated by $F_1$. In the second stage, namely refinement stage, one examines the pair identified from the first stage more closely and picks one sequence to be the final sequence. The problem under this setup belongs to a class of multiple stopping time problems. In particular, it is an ordered two concatenated Markov stopping time problem. We obtain the optimal solution using the tools from the multiple stopping time theory. Numerical simulation results show that this search strategy can significantly reduce the searching time, especially when $F_{1}$ is rare.
\end{abstract}

\section{Introduction} \label{sec:intro}

The quickest search over multiple sequences problem, a generalization of the classical sequential hypothesis testing problem~\cite{Wald:AMS:45}, was originally proposed in a recent paper~\cite{Lai:TIT:11}. In particular, the author considered a case that multiple sequences are available. For each individual sequence, it may either be generated by distribution $F_{0}$ or $F_{1}$, and its distribution is independent of all other sequences. A sensor can take observations from these sequences, and the goal is to find a sequence which is generated by $F_{1}$ as quickly as possible under an error probability constraint. Assuming that the sensor can take one observation from a {\it single} sequence at a time, \cite{Lai:TIT:11} showed that the cumulative sum (CUSUM) test is optimal. This quickest search problem has applications in various field such as cognitive radio and database search. The sample complexity of a such search problem is analyzed in~\cite{Malloy:TIT:12}.~\cite{Bayraktar:12} studies the search problem over continuous time Brownian channels. The problem of recovering more than one sequence generated from $F_1$ is considered in~\cite{Tajer:ALL:12}.

In this paper, we propose a new search approach: search with mixed observations. This search strategy consists of two stages. In the first stage, namely the scanning stage, the sensor takes observations that are linear combinations of samples from a pair of different sequences. In certain applications, such as cognitive radios, it is easy to obtain an observation that is a linear combination of signals from different sequences. The purpose of this stage is to scan through sequences generated by $F_0$ and quickly identify a pair of sequences among which at least one of them is generated by $F_{1}$. In particular, if the sensor believes that both of the sequences that generate the observation are from $F_{0}$, then it discards these two sequences and switches to observe two new sequences. Otherwise, the sensor stops the scanning stage and enters the refinement stage. In the refinement stage, the sensor examines one of the two candidate sequences identified in the scanning stage carefully, and makes a final decision on which one of the two sequences is generated by $F_{1}$. Hence, in the refinement stage, no mixing is used anymore.

With this mixed observation strategy, our goal is to minimize a linear combination of the searching delay and the error probability. Toward this goal, we need to optimize over four decision rules: 1) the stopping time for the scanning stage $\tau_{1}$, which determines when one should stop the scanning stage and enter the refinement stage; 2) the sequence switching rule in the scanning stage $\phiv$, which determines when one should switch to new sequences for scanning; 3) the stopping time for refinement stage $\tau_{2}$, which determines when one should stop the whole search process; and 4) the final decision rule in the refinement stage $\delta$, which determines which sequence will be claimed to be generated from $F_1$. This two stage search problem can be converted to an optimal multiple stopping time problem~\cite{Kobylanski:AAP:11}. In particular, we show that this problem can be converted into an ordered two concatenated Markov stopping time problems. Using the optimal multiple stopping time theory~\cite{Kobylanski:AAP:11}, we derive the optimal strategy for this search problem. We show that the optimal solutions of $\tau_{1}$ and $\phiv$ turn out to be region rules. The optimal solution for $\tau_{2}$ is the time when the error probability cost less than the future cost, and the optimal decision rule $\delta$ is to pick the sequence with a larger posterior probability of being generated by $F_{1}$.

The motivation to propose this mixed observation searching strategy is to improve the search efficiency when the presence of $F_{1}$ is rare. If most of the sequences are generating by $F_{0}$, then the sensor can scan through and discard the sequences more quickly by this mix strategy. Our numerical results show that our strategy can significantly reduce the search delay when $F_{1}$ is rare. In some sense, our strategy has a similar flavor with that of the group testing~\cite{Dorfman:AmS:43} and compressive sensing~\cite{Donoho:TIT:06} in which linear combinations of signals are observed.

The remainder of the paper is organized as follows. The mathematical model is given in Section~\ref{sec:model}. Section~\ref{sec:solution} presents the optimal solution to this quickest search problem. Numerical examples are given in Section~\ref{sec:numerical}. Finally, Section~\ref{sec:conclusion} offers concluding remarks. Due to space limitations, we omit the details of the proofs.

\section{Model} \label{sec:model}
We consider $N$ sequences $\{Y_{k}^{i};k=1,2,\cdots\}, i=1,\cdots,N$, where for each $i$, $\{Y_{k}^{i};k=1,2,\cdots\}$ are i.i.d. observations taking values in a set $\Omega$ endowed with a $\sigma$-field $\mathcal{F}$ of events, that obey one of the two hypotheses:
\begin{eqnarray}
&&H_{0}:\quad Y_{k}^{i}\sim F_{0},\quad k=1,2,\cdots\no\\
\text{versus}&&\no\\
&&H_{1}:\quad Y_{k}^{i}\sim F_{1},\quad k=1,2,\cdots, \no
\end{eqnarray}
where $F_0$ and $F_1$ are two distinct, but equivalent, distributions on $(\Omega,\mathcal{F})$. We use $f_0$ and $f_1$ to denote densities of $F_0$ and $F_1$, respectively, with respect to some common dominating measure. The sequences for different values of $i$ are independent. Moreover, whether the $i^{th}$ sequence $\{Y_{k}^{i};k=1,2,\cdots\}$ is generated by $F_0$ or $F_1$ is independent of all other sequences. Here, we assume that for each
$i$, hypothesis $H_1$ occurs with prior probability $\pi$ and $H_0$ with prior probability $1-\pi$. The goal of the quickest search is to locate a sequence that is generated from $F_1$ quickly and accurately.

The search strategy has two stages, namely the scanning stage and the refinement stage. In the scanning stage, the sensor observes a linear combination of samples from two sequences, and decides whether at least one of the observed sequences is generated by $F_{1}$. If the sensor has enough confidence on that one of the observed sequences is generated by $F_{1}$, it enters the refinement stage, in which the sensor examines these two sequences identified in scanning stage carefully, and decides which sequence is generate by $F_{1}$.

Specifically, in the scanning stage, at each time slot $k$, the sensor picks two sequences $s_k^{a}$ and $s_k^{b}$, and observes a linear combination of samples from these two sequences:
\begin{eqnarray}
Z_k=a_1Y_{k}^{s_k^{a}}+a_2Y_{k}^{s_k^{b}}.
\end{eqnarray}
In this paper, we set $a_1=a_2=1$, which might not the optimal choice. We will study the optimal choice of these two parameters in our future work. Since each sequence has two possible pdfs, $Z_k$ has three possible pdfs:
\begin{enumerate}
\item $f_{0,0}\triangleq f_0*f_0$, which happens when both sequences $s_k^{a}$ and $s_k^{b}$ are generated from $f_0$. Here $*$ denotes convolution. The prior probability of this occurring is $p^{0,0}_0=(1-\pi)^2$;

\item $f_{m}\triangleq f_0*f_1$, which happens when one of these two sequences is generated from $f_0$ and the other one is generated from $f_1$. The prior probability of this occurring is $p^{mix}_0=2\pi(1-\pi)$;

\item $f_{1,1}\triangleq f_1*f_1$, which happens when both sequences are generated from $f_1$. The prior probability of this occurring is $p^{1,1}_0=\pi^2$.
\end{enumerate}

After taking observation $Z_{k}$, the sensor needs to make one of the following three decisions: 1) to continue the scanning stage and to take one more observation from these two currently observing sequences; or 2) to continue the scanning stage but to take observation from two other sequences, that is, the sensor discards the currently observing sequences and switches to observe two new sequences; or 3) to stop the scanning stage and to enter the refinement stage to further examine these two candidate sequences. Hence, there are two decisions in the scanning stage: the stopping time $\tau_{1}$, at which the sensor stops the scanning stage and enters the refinement stage, and the sequences switching rule $\phiv=(\phi_1,\phi_2,\cdots)$, based on which the sensor abandons the currently observing  sequences and switches to observe new sequences. Here, the element $\phi_{k} \in \{0, 1\}$ denotes the sequence switching status at time slot $k$. Specifically, if $\phi_{k}=1$, the sensor switches to new sequences, while if $\phi_{k}=0$, the sensor keeps observing the current two sequences. Here, the stopping time $\tau_{1}$ is adapted to the filtration $\mathcal{F}_k=\sigma(Z_1,\cdots,Z_k)$, and the switching rule $\phi_{k}$ is a measurable function of $\mathcal{F}_k$. 

In the refinement stage, we examine the two candidate sequences more closely. Each observation taken during the refinement stage will come from one sequence. Hence, at this stage, no mixing is used anymore. The observation sequence in the refinement stage is denoted as $\{X_{j}, j=1, 2, \ldots\}$. Clearly, at the beginning of the refinement stage, i.e. $j=1$, there is no difference between these two candidates $s_{\tau_1}^{a}$ and $s_{\tau_1}^{b}$, and hence the sensor simply picks one $s_{\tau_1}^{a}$ to observe:
\begin{eqnarray}
X_j=Y_{\tau_1+j}^{s_{\tau_1}^{a}}.
\end{eqnarray}
After taking an observation $X_{j}$, the sensor needs to decide whether or not to stop the refinement stage, and if so, the sensor should pick one sequence from $s_{\tau_1}^{a}$ and $s_{\tau_1}^{b}$, and claim that it is generated from $f_1$. Again, there are two decisions in this stage: the stopping time $\tau_{2}$, at which the sensor decides to stop the refinement stage, and the terminal decision rule $\delta$ that determines which sequence to be claimed as being generated by $f_{1}$. $\tau_{2}$ is adapted to the filtration $\mathcal{G}_j=\sigma(Z_1,\cdots,Z_{\tau_1},X_1,\cdots,X_j)$.

Two performance metrics are of interest: the total time spent on the search process $\tau_1+\tau_2$ and the error probability such that the picked sequence is generated from $f_0$. Clearly, if we spend more time on the search, the error probability will be reduced. We aim to minimize a cost function which is a linear combination of these two quantities. Hence, our goal is to design $\tau_1$, $\phiv$, $\tau_2$ and $\delta$ to solve the following optimization problem:
\begin{eqnarray}\label{eq:cost}
\inf\limits_{\tau_1,\phiv,\tau_2,\delta}\left\{c\mathbb{E}[\tau_1+\tau_2]+\text{Pr}\left(H^{\delta}=H_0\right)\right\}.
\end{eqnarray}
We note that there are two inter-related stopping times involved in the problem.

\section{Solution} \label{sec:solution}
In this section, we discuss the optimal solution for the proposed sequential search problem. We first introduce some important statistics used in the optimal solution.

For the scanning stage, after taking $k$ observations, we define the following posterior probabilities:
\begin{eqnarray}
&&p^{1,1}_k = \text{Pr}\left\{ Y_{k}^{s_k^a} \sim f_1,  Y_{k}^{s_k^b} \sim f_1 \Big| \mathcal{F}_k \right\}, \no \\
&&p^{mix}_k = \text{Pr}\left\{ Y_{k}^{s_k^a} \sim f_0,  Y_{k}^{s_k^b} \sim f_1 \text{ or } \right. \no \\
&&\quad \quad \quad \quad \quad \left. Y_{k}^{s_k^a} \sim f_1,  Y_{k}^{s_k^b} \sim f_0 \Big| \mathcal{F}_k \right\}, \no \\
&&p^{0,0}_k = \text{Pr}\left\{ Y_{k}^{s_k^a} \sim f_0,  Y_{k}^{s_k^b} \sim f_0 \Big| \mathcal{F}_k \right\}. \no
\end{eqnarray}
As discussed in Section \ref{sec:model}, at the beginning of the scanning stage we have $p^{1,1}_0 = \pi^2$, $p^{mix}_0 = 2\pi(1-\pi)$ and $p^{0,0}_0 = (1-\pi)^2$.

It is easy to check that these posterior probabilities can be updated as follows:
\begin{eqnarray}
&&\hspace{-8mm}p^{1,1}_{k+1}=\frac{p^{1,1}_{k}f_{1,1}(Z_{k+1})}{f_{Z,k}(Z_{k+1})}\mathbf{1}_{\{\phi_k=0\}}+\frac{p^{1,1}_{0}f_{1,1}(Z_{k+1})}{f_{Z,0}(Z_{k+1})}\mathbf{1}_{\{\phi_k=1\}},\no\\
&&\hspace{-8mm}p^{mix}_{k+1}=\frac{p^{mix}_{k}f_{m}(Z_{k+1})}{f_{Z,k}(Z_{k+1})}\mathbf{1}_{\{\phi_k=0\}}+\frac{p^{mix}_{0}f_{m}(Z_{k+1})}{f_{Z,0}(Z_{k+1})}\mathbf{1}_{\{\phi_k=1\}},\no\\
&&\hspace{-8mm}p^{0,0}_{k+1}=1-p^{1,1}_{k+1}-p^{mix}_{k+1}, \no
\end{eqnarray}
where $\mathbf{1}$ is the indicator function, $f_{Z,k}(z_{k+1})$ and $f_{Z,0}(z_{k+1})$ are defined as
\begin{eqnarray}
&&\hspace{-10mm}f_{Z,k}(z_{k+1}) \triangleq \no\\
&& p^{0,0}_{k}f_{0,0}(z_{k+1})+p^{mix}_kf_{m}(z_{k+1})+p^{1,1}_kf_{1,1}(z_{k+1}),\no\\
&&\hspace{-10mm}f_{Z,0}(z_{k+1}) \triangleq \no\\
&& p^{0,0}_{0}f_{0,0}(z_{k+1})+p^{mix}_0f_{m}(z_{k+1})+p^{1,1}_0f_{1,1}(z_{k+1}).\no
\end{eqnarray}

For the refinement stage, after taking $j$ observations, we define
\begin{eqnarray}
r^{1,1}_{j} = \text{Pr}\left\{Y_{\tau_{1}+j}^{s_{\tau_1}^a} \sim f_1, Y_{\tau_{1}+j}^{s_{\tau_1}^b} \sim f_1 \Big|\mathcal{G}_j \right\}, \no \\
r^{1,0}_{j} = \text{Pr}\left\{Y_{\tau_{1}+j}^{s_{\tau_1}^a} \sim f_1, Y_{\tau_{1}+j}^{s_{\tau_1}^b} \sim f_0 \Big|\mathcal{G}_j \right\}, \no \\
r^{0,1}_{j} = \text{Pr}\left\{Y_{\tau_{1}+j}^{s_{\tau_1}^a} \sim f_0, Y_{\tau_{1}+j}^{s_{\tau_1}^b} \sim f_1 \Big|\mathcal{G}_j \right\}, \no \\
r^{0,0}_{j} = \text{Pr}\left\{Y_{\tau_{1}+j}^{s_{\tau_1}^a} \sim f_0, Y_{\tau_{1}+j}^{s_{\tau_1}^b} \sim f_0 \Big|\mathcal{G}_j \right\}. \no
\end{eqnarray}
At the beginning of the refinement stage, we have
\begin{eqnarray}
&& r^{1,1}_{0} = p_{\tau_{1}}^{1,1}, \no\\
&& r^{1,0}_{0} = r^{0,1}_{0} = p_{\tau_{1}}^{mix}/2. \no
\end{eqnarray}
It is easy to verify that these statistics can be updated using
\begin{eqnarray}
&&\hspace{-6mm} r^{1,1}_{j+1} = \frac{f_{1}(X_{j+1}) r^{1,1}_{j}}{f_{1}(X_{j+1})(r^{1,1}_{j} + r^{1,0}_{j}) + f_{0}(X_{j+1}) (r^{0,1}_{j}+r^{0,0}_{j})}, \no\\
&&\hspace{-6mm} r^{1,0}_{j+1} = \frac{f_{1}(X_{j+1}) r^{1,0}_{j}}{f_{1}(X_{j+1})(r^{1,1}_{j} + r^{1,0}_{j}) + f_{0}(X_{j+1}) (r^{0,1}_{j}+r^{0,0}_{j})}, \no\\
&&\hspace{-6mm} r^{0,1}_{j+1} = \frac{f_{0}(X_{j+1}) r^{0,1}_{j}}{f_{1}(X_{j+1})(r^{1,1}_{j} + r^{1,0}_{j}) + f_{0}(X_{j+1}) (r^{0,1}_{j}+r^{0,0}_{j})}, \no\\
&&\hspace{-6mm} r^{0,0}_{j+1} = 1 - r^{1,1}_{j+1} - r^{1,0}_{j+1} - r^{0,1}_{j+1}. \no
\end{eqnarray}
For the brevity of notation, we further define the following two statistics
$\pi^{s_{\tau_1}^a}_j=\text{Pr}\left\{Y_{k}^{s_k^a} \sim f_1 \Big| \mathcal{G}_j \right\}, $
$\pi^{s_{\tau_1}^b}_j=\text{Pr}\left\{Y_{k}^{s_k^b} \sim f_1 \Big| \mathcal{G}_j \right\}. $
By definition, it is easy to verify that
$\pi^{s_{\tau_1}^a}_{j} = r^{1,1}_{j} + r^{1,0}_{j}, $
$ \pi^{s_{\tau_1}^b}_{j} = r^{1,1}_{j} + r^{0,1}_{j}.$

In the following, we solve the multiple stopping time optimization problem \eqref{eq:cost}. We solve this problem by decomposing it into two single stopping time problems, which are corresponding to the refinement stage and scanning stage respectively. First, we have following lemma. 
\begin{lem} \label{lem:error_prob}
For any $\tau_1,\phiv$ and $\tau_2$, the optimal decision rule is given as
\begin{eqnarray}
\delta^{*}=\left\{\begin{array}{ll}s_{\tau_1}^a & \text{ if } \pi^{s_{\tau_1}^a}_{\tau_2} > \pi^{s_{\tau_1}^b}_{\tau_2} \\
s_{\tau_1}^b &\text{ if } \pi^{s_{\tau_1}^a}_{\tau_2} \leq \pi^{s_{\tau_1}^b}_{\tau_2} \end{array}\right. ,
\end{eqnarray}
and the corresponding cost is given as
\begin{eqnarray}
\inf\limits_{\delta} \text{Pr}\left(H^{\delta}=H_0\right) = \mathbb{E}\left[1-\max\left\{\pi^{s_{\tau_1}^a}_{\tau_2},\pi^{s_{\tau_1}^b}_{\tau_2}\right\}\right].
\end{eqnarray}
\end{lem}

This lemma converts the cost of the error probability to a function of $\pi_{j}^{s_{\tau_1}^a}$ and $\pi_{j}^{s_{\tau_1}^b}$, which is a function of the refinement stage statistics $\mathbf{r}_{j} = \left[r_{j}^{1,1}, r_{j}^{0,1}, r_{j}^{1,0}\right]$. In the following, we first consider the refinement stage optimization problem for any given $\tau_1$ and $\phiv$:
%
\begin{eqnarray}
\inf_{\tau_2,\delta}\mathbb{E}\left[c \tau_2 + \text{Pr}\left(H^{\delta}=H_0\right) | \mathcal{F}_{\tau_1} \right]. \label{eq:cost_refinement}
\end{eqnarray}
The optimal stopping time for $\tau_{2}$ is given as:
\begin{lem} \label{lem:refine}
For any given $\tau_1$ and $\phiv$, the optimal stopping time $\tau_{2}$ is given as
\begin{eqnarray}\
\tau_{2}^{*} &=& \inf\left\{ j \geq 0: 1 - \max\left\{\pi_{j}^{s_{\tau_{1}}^{a}}, \pi_{j}^{s_{\tau_{1}}^{b}} \right\} \leq \right.\no\\
&&\left.\hspace{15mm}c+ \mathbb{E}\left[ V_{r}(\mathbf{r}_{j+1}) \Big| \mathbf{r}_{j} \right] \right\}, \no
\end{eqnarray}
in which $V_{r}(\mathbf{r}_{j})$ is a function that satisfies the following operator:
\begin{eqnarray}
V_{r}(\mathbf{r}_{j}) &=& \min \left\{ 1 - \max\left\{\pi_{j}^{s_{\tau_{1}}^{a}}, \pi_{j}^{s_{\tau_{1}}^{b}} \right\}, \right. \no\\
&&\left.\hspace{10mm} c + \mathbb{E}\left[ V_{r}(\mathbf{r}_{j+1}) \Big| \mathbf{r}_{j} \right] \right\}. \no
\end{eqnarray}
\end{lem}
We note that the form of $V_{r}(\mathbf{r}_{j})$ can be obtained via an iterative procedure offline~\cite{Poor:Book:08}. This lemma indicates that the optimal strategies in the refinement stage are related to $\tau_1, \phiv$ only through $p^{1,1}_{\tau_{1}},p^{mix}_{\tau_{1}}$. Hence, the minimal cost of the refinement stage is a function of only $p^{1,1}_{\tau_{1}},p^{mix}_{\tau_{1}}$:
\begin{eqnarray}
g\left(p^{1,1}_{\tau_{1}},p^{mix}_{\tau_{1}}\right) &\triangleq& \inf_{\tau_2,\delta}\mathbb{E}\left[c \tau_2 + \text{Pr}\left(H^{\delta}=H_0\right) | \mathcal{F}_{\tau_1} \right]. \no\\
&=& V_{r}(p^{1,1}_{\tau_{1}}, p^{mix}_{\tau_{1}}/2, p^{mix}_{\tau_{1}}/2). \no
\end{eqnarray}
It is defined over the domain
\begin{eqnarray}
&&\mathcal{P} = \left\{ \left(p^{1,1},p^{mix}\right): 0 \leq p^{1,1} \leq 1, 0 \leq p^{mix} \leq 1,\right. \no\\
&&\quad \quad \quad \quad  \quad \quad \quad \quad \quad \left. 0 \leq p^{1,1}+p^{mix} \leq 1 \right\}.\no
\end{eqnarray}

\begin{prop}
$g\left(p^{1,1},p^{mix}\right)$ is a concave function over $\mathcal{P}$
with $g(1,0) = 0$ and $g(0,0) = 1$.
\end{prop}

As the result, the original problem \eqref{eq:cost} can be converted into an equivalent problem with respect to only $\tau_{1}$ and $\phiv$. Since
\begin{eqnarray}
&&\hspace{-10mm}\inf\limits_{\tau_1,\phiv,\tau_2,\delta}\mathbb{E}\left[c(\tau_{1} + \tau_{2}) + \text{Pr}\left(H^{\delta}=H_0\right)\right] \no \\
&=& \inf\limits_{\tau_1,\phiv,\tau_2,\delta}\mathbb{E}\left[c\tau_{1} + \mathbb{E} \left[c \tau_{2} + \text{Pr}\left(H^{\delta}=H_0\right)| \mathcal{F}_{\tau_1} \right]\right] \no\\
&\geq& \inf\limits_{\tau_1,\phiv} \mathbb{E}\left[c\tau_{1} + g\left(p^{1,1}_{\tau_{1}},p^{mix}_{\tau_{1}}\right)\right], \no
\end{eqnarray}
the equality holds if using $\tau_{2}^{*}$ and $\delta^{*}$ specified in Lemma \ref{lem:refine} and \ref{lem:error_prob}, respectively. Therefore, ~\eqref{eq:cost} is equivalent to
\begin{eqnarray}
\inf\limits_{\tau_1,\phiv}\mathbb{E}\left[c\tau_1 + g\left(p^{1,1}_{\tau_{1}},p^{mix}_{\tau_{1}}\right)\right]. \label{eq:cost_scan}
\end{eqnarray}

\begin{lem}
The optimal stopping rule for the scanning stage is given as
\begin{eqnarray}
\tau_{1}^{*} = \inf\left\{ k\geq 0: g\left(p_{k}^{1, 1}, p_{k}^{mix}\right) = V_{s}\left(p_{k}^{1, 1}, p_{k}^{mix}\right) \right\}
\end{eqnarray}
and the optimal switching rule is given as
\begin{eqnarray}
\phi_k^{*} = \left\{\begin{array}{ll} 0 & \text{ if } A_{c}\left(p_{k}^{1,1}, p_{k}^{mix}\right) \leq A_{s}\\
1 &\text{ otherwise } \end{array}\right.,
\end{eqnarray}
in which, $V_s(\cdot)$ is a function that satisfies the following operator
\begin{eqnarray}
V_{s}\left(p_{k}^{1, 1}, p_{k}^{mix}\right) &=& \min \left\{ g\left(p_{k}^{1, 1}, p_{k}^{mix}\right), \right. \no \\
&&\hspace{-4mm} \left. c + \min \left\{ A_{c}\left(p_{k}^{1, 1}, p_{k}^{mix}\right), A_{s} \right\} \right\} \no
\end{eqnarray}
with
\begin{eqnarray}
&&\hspace{-7mm} A_{c}\left(p_{k}^{1, 1}, p_{k}^{mix}\right) = \mathbb{E}\left[V_{s}\left(p_{k+1}^{1, 1}, p_{k+1}^{mix}\right)\Big|p_{k}^{1, 1}, p_{k}^{mix}, \phi_{k}=0 \right], \no\\
&&\hspace{-7mm} A_{s} = \mathbb{E}\left[V_{s}\left(p_{k+1}^{1, 1}, p_{k+1}^{mix}\right)\Big|p_{k}^{1, 1}, p_{k}^{mix}, \phi_{k}=1 \right]. \no
\end{eqnarray}

\end{lem}
\begin{rmk}
Same as above, all the functions involved in this lemma can be computed offline.
\end{rmk}
\begin{rmk}
One can show that $A_{s} = A_{c}\left(p_{0}^{1, 1}, p_{0}^{mix}\right)$, hence it is a constant between 0 and 1. For this reason, we denote it as $A_{s}$ rather than $A_{s}\left(p_{k}^{1, 1}, p_{k}^{mix}\right)$ in the above lemma.
\end{rmk}

The optimal solutions of $\tau_{1}^{*}$ and $\phi_{k}^{*}$ can be further simplified using the following proposition.
\begin{prop}
1) $V_{s}\left(p^{1, 1}, p^{mix}\right)$ is a concave function over domain $\mathcal{P}$, and $0 \leq V_{s}\left(p^{1, 1}, p^{mix}\right) \leq 1$. \\ 
2) $A_{c}\left(p^{1,1}, p^{mix}\right)$ is a concave function over $\mathcal{P}$, and $0 \leq A_{c}\left(p^{1, 1}, p^{mix}\right) \leq 1$. 
\end{prop}

Since both $V_{s}\left(p^{1, 1}, p^{mix}\right)$ and $g\left(p^{1, 1}, p^{mix}\right)$ are concave functions over $\mathcal{P}$, $V_{s}\left(p^{1, 1}, p^{mix}\right) \leq g\left(p^{1, 1}, p^{mix}\right)$ over $\mathcal{P}$, and $V_{s}(1, 0) = g(1,0) = 0$, there must exist some region, denoted as $R_{\tau}$, on which these two concave surfaces are equal to each other. Hence, the optimal stopping time $\tau_{1}^{*}$ can be described as the first hitting time of the process $\left(p_{k}^{1, 1}, p_{k}^{mix}\right)$ to region $R_{\tau}$. Similarly, $A_{c}$ is a concave surface and $A_{s}$ is a constant plane with $A_{s} = A_{c}(p^{1,1}_{0}, p^{mix}_{0})$. Hence, $\mathcal{P}$ can be divided into two connected regions $R_{\phi}$ and $\mathcal{P}\backslash R_{\phi}$, where $R_{\phi} \triangleq \left\{\left(p^{1,1}, p^{mix}\right): A_{c}(p^{1,1}, p^{mix}) \leq A_{s} \right\}$. Hence, the sensor switches to new sequences at time slot $k$ if $\left(p^{1,1}_{k}, p^{mix}_{k}\right)$ is in $R_{\phi}$. Hence, we have the following lemma.

\begin{lem}\label{lem:scan}
There exist two regions, $R_{\tau} \subset \mathcal{P}$ and $R_{\phi} \subset \mathcal{P} $, such that
\begin{eqnarray}
\tau_1^{*} = \min\left\{k \geq 0: \left(\pi_{k}^{1,1}, \pi_{k}^{mix}\right) \in R_{\tau} \right\},
\end{eqnarray}
and
\begin{eqnarray}
\phi_k^{*} =\left\{\begin{array}{ll} 1 & \text{ if } (\pi_{k}^{1,1}, \pi_{k}^{mix}) \in R_{\phi}\\
0 &\text{ otherwise } \end{array}\right..
\end{eqnarray}
\end{lem}

\section{Simulation} \label{sec:numerical}
In this section, we give some numerical examples to illustrate the analytical results of the previous sections. In these numerical examples, we assume $f_{0} \sim \mathcal{N}(0, \sigma^2)$ and $f_{1} \sim \mathcal{N}(0, P+\sigma^2)$. The signal-to-noise ratio is defined as $SNR = 10 \log P/\sigma^2$.

\begin{figure}[thb]
\centering
\includegraphics[width=0.3 \textwidth]{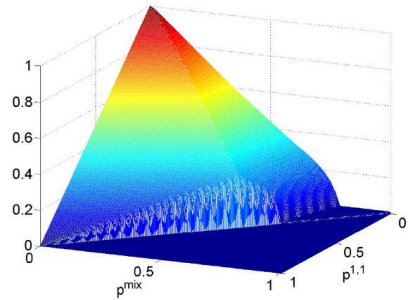}
\caption{An illustration of $g(p^{1,1}, p^{mix})$}
\label{fig:cost_refine}
\end{figure}

In the first scenario, we illustrate the cost function of the refinement procedure $g\left(p^{1,1}, p^{mix}\right)$. In this simulation, we choose $\pi = 0.05$, $c=0.01$, $\sigma^2 = 1$ and $SNR = 3dB$. The simulation result is shown in Figure \ref{fig:cost_refine}. This simulation confirms our analysis that $g\left(p^{1,1}, p^{mix}\right)$ is a concave function within $[0, 1]$ over $\mathcal{P}$. We also notice that $g(1,0) = 0$ and $g(0, 0)=1$, this is reasonable since if the sensor knows both $s_{\tau_{1}}^{a}$ and $s_{\tau_{1}}^{b}$ are generated by $f_{1}$, which corresponding to $p^{1,1}_{\tau_{1}} = 1$ and $p^{mix}_{\tau_{1}} = 0$, the sensor can make a decision on either of sequences without taking any observation and making any error, hence the cost on refinement stage is $0$. Similarly, if the sensor knows neither $s_{\tau_{1}}^{a}$ nor $s_{\tau_{1}}^{b}$ is generated by $f_{1}$, that is $p^{1,1}_{\tau_{1}} = 0$ and $p^{mix}_{\tau_{1}} = 0$, no matter what decision is made, the cost of error would be $1$. We also notice that in the area close to $p^{1,1}_{\tau_{1}} = 0, p^{mix}_{\tau_{1}} = 1$, which indicates the sensor is quite sure that one of sequences is generated by $f_{1}$, the cost is small. This is because the sensor can significantly reduce the cost of decision error by taking a few observations.

\begin{figure}[thb]
\centering
\includegraphics[width=0.3 \textwidth]{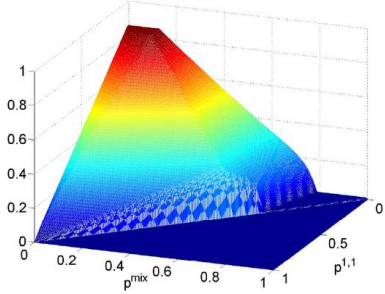}
\caption{An illustration of $V_{s}(p^{1,1}, p^{mix})$}
\label{fig:cost_total}
\end{figure}

\begin{figure}[thb]
\centering
\includegraphics[width=0.3 \textwidth]{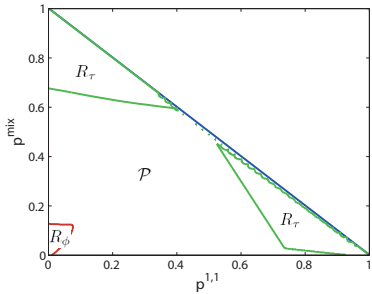}
\caption{The optimal stopping and switching regions}
\label{fig:regions}
\end{figure}

In the second scenario, we illustrate the overall cost function $V_{s}\left(p^{1,1}, p^{mix}\right)$ using the same simulation parameters. The simulation result is shown in Figure \ref{fig:cost_total}. This simulation confirms that $V_{s}\left(p^{1,1}, p^{mix}\right)$ is also a concave function over $\mathcal{P}$. Moveover, this function is flat on top since it is upper bounded by constant $c+A_{s}$. This flat area corresponds to $R_{\phi}$, hence if $\left(p^{1,1}_{k}, p^{mix}_{k}\right)$ enters this region, the sensor would switch to scan new sequences at time slot $k$. Similarly, the cost function is also upper bounded by $g\left(p^{1,1}, p^{mix}\right)$, which is shown in Figure \ref{fig:cost_refine}. On the region, $R_{\tau}$, that these two surfaces overlap each other, the sensor would stop the scanning stage and enter the refinement stage. The location of $R_{\phi}$ and $R_{\tau}$ is illustrated in Figure \ref{fig:regions}. In this figure, the left-lower half below the blue line is the domain $\mathcal{P}$. The region circled by the red line is the sequence switching region $R_{\phi}$, and the region circled by green is the scanning stop region $R_{\tau}$. In this simulation, $R_{\tau}$ are two separate regions located around $(0, 1)$ and $(1,0)$ respectively, which means the sensor will enter the refinement stage as long as it has enough confidence on that at least one of the observed sequences is generated by $F_{1}$. $R_{\tau}$ and $R_{\phi}$ can be calculated off line.

\begin{figure}[thb]
\centering
\includegraphics[width=0.3 \textwidth]{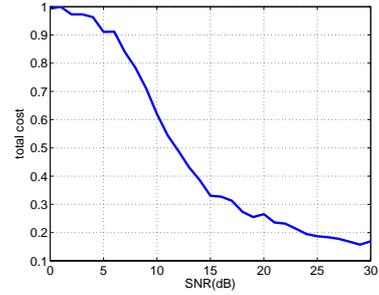}
\caption{The total cost vs. SNR under $c = 0.01$ and $\pi =0.05$}
\label{fig:cost_vs_snr}
\end{figure}

In the next scenario, we illustrate the relationship between total cost and SNR. The total cost consists of two parts: the cost of searching delay 
and the cost of error probability 
. In the simulation, we choose $\pi=0.05$, $c=0.01$ and $\sigma^{2} = 1$. The simulation result is shown in Figure \ref{fig:cost_vs_snr}. From the figure we can see that the curve follows a decreasing trend. This is consistent with our intuition, that is, the higher SNR is, the easier it is to distinguish between these two hypotheses.

\begin{figure}[thb]
\centering
\includegraphics[width=0.3 \textwidth]{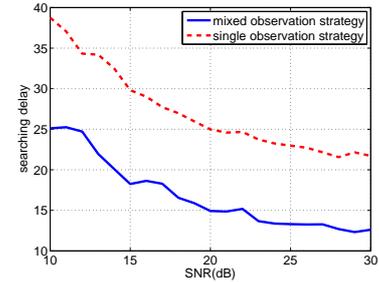}
\caption{The comparison of delays for different searching strategies}
\label{fig:comparsion}
\end{figure}

In the last scenario, we compare the proposed strategy with the strategy proposed by \cite{Lai:TIT:11}, which does not allow observation mixing and is referred as the single observation strategy in the sequel. We compare the search delays of these two strategies by keeping the error probabilities to a same level. In this simulation, we choose $\pi = 0.05$. Figure~\ref{fig:comparsion} shows the simulation result. In this figure, the blue solid line is the search delay induced by the mix observation strategy, and the red dash line is the searching delay by the single observation strategy. As we can see, the proposed mix observation strategy saves about $40\%$ of the search time since discarding two sequences together is more efficient than discarding sequences one by one.

\section{Conclusion} \label{sec:conclusion}
In this paper, we have proposed a new search strategy for the quickest search over multiple sequences problem. We have formulated this problem as an optimal multiple stopping time problem. 
We have solved this problem by decomposing the problem into an ordered two concatenated Markov stopping time problem. Our simulation result shows that when $H_{1}$ rarely occurs, the proposed strategy can significantly reduce the search delay. In terms of the future work, it is interesting and important to analytically quantify the performance gain.
\bibliographystyle{ieeetr}
\bibliography{macros,sensornetwork,secrecy}

\end{document}